\documentclass[manuscript]{acmart}
\usepackage{multirow}
\AtBeginDocument{%
  \providecommand\BibTeX{{%
    \normalfont B\kern-0.5em{\scshape i\kern-0.25em b}\kern-0.8em\TeX}}}

\begin{document}

\title{Tensor-based Collaborative Filtering With Smooth Ratings Scale}


\author{Nikita Marin}
\affiliation{%
  \institution{Skolkovo Institute of Science and Technology}
  \city{Moscow}
  \country{Russia}}
\email{marin.nikita@mail.ru}

\author{Elizaveta Makhneva}
\affiliation{%
  \institution{Skolkovo Institute of Science and Technology}
  \city{Moscow}
  \country{Russia}}
\email{elizavetamakhneva@mail.ru}

\author{Maria Lysyuk}
\affiliation{%
  \institution{Skolkovo Institute of Science and Technology}
  \city{Moscow}
  \country{Russia}}
\email{mlysyuk@nes.ru}

\author{Vladimir Chernyy}
\affiliation{%
  \institution{Skolkovo Institute of Science and Technology}
  \city{Moscow}
  \country{Russia}}
\email{Vladimir.Chernyy@skoltech.ru}

\author{Ivan Oseledets}
\additionalaffiliation{%
  \institution{AI Research Institute, Moscow, Russia}}
\affiliation{%
  \institution{Skolkovo Institute of Science and Technology}
  \city{Moscow}
  \country{Russia}}
\email{I.Oseledets@skoltech.ru}

\author{Evgeny Frolov}
\affiliation{%
  \institution{Skolkovo Institute of Science and Technology}
  \city{Moscow}
  \country{Russia}}
\email{Evgeny.Frolov@skoltech.ru}

\renewcommand{\shortauthors}{Marin and Makhneva, et al.}

\begin{abstract}
  Conventional collaborative filtering techniques don't take into consideration the effect of discrepancy in users' rating perception. Some users may rarely give 5 stars to items while others almost always assign 5 stars to the chosen item. Even if they had experience with the same items this systematic discrepancy in their evaluation style will lead to the systematic errors in the ability of recommender system to effectively extract right patterns from data. To mitigate this problem we introduce the ratings' similarity matrix which represents the dependency between different values of ratings on the population level. Hence, if on average the correlations between ratings exist, it is possible to improve the quality of proposed recommendations by off-setting the effect of either shifted down or shifted up users' rates.
\end{abstract}




\maketitle

\section{Introduction}

Recommendations reliability significantly depends on the unbiasedness of feedback provided by users. In many contemporary recommendation models, it is often implicitly assumed that user feedback explicitly expressed in the form of e.g. ratings represents an accurate picture of a perceived quality of products and allows to unambiguously discern subjective preferences of individuals. On the other hand, there is a widespread concern regarding noise and inconsistencies in user ratings stemming from a wide range of possible reasons.

For instance, in psychology, there is a well-known assimilation/contrast effect \cite{sherif1961social}. Users tend to conform with historical ratings if the perceived product quality is not far from its rating (assimilation). On the contrary, if the product quality is too far from the historical ratings they start deviating from these benchmarks (contrast). In this regard, feedback loops -- an inevitable attribute of recommender systems -- are likely to amplify such effects to a greater extent. For example, if an online system exposes average product ratings to users, it may steer their provided individual feedback, which in turn will lead to even more inaccurate average ratings estimation and spiral the loop further. 

Another inconsistency may stem from the evolution of provided user feedback in time. \citet{cosley2003seeing} conducted an experiment with 212 participants who were asked to re-rate 40 movies they had already seen and rated in the center of the 1-5 scale (i.e., from 2 to 4). The results showed that the participants were consistent with their initial opinion only in 60\% of the cases. As a follow-up, \citet{amatriain2009like} learnt the structure of these differences. They found that the probability of inconsistencies was the largest for 2- and 3-star ratings at the 5-star scale. That is, given two consequent trials $x_1$ and $x_2$ from the same individual, the most frequent change pattern was either switching from 3 stars at $x_1$ to 2 stars at $x_2$, or replacing 4 stars at $x_1$ with 3 stars at $x_2$. Other combinations with even more pronounced shifts were also present, albeit less frequently. In general, it turned out that users were likely to change their opinion about the same movies within just a few weeks.

The striking conclusion that can be made from these observations is not only that different users may have different scales of a perceived quality of products, but even the same user's scale may shift in time and also depend on some external factors and context. The examples above expose an intricate subjective nature of explicit user feedback, which in turn may have a significant impact on real-world personalization services. An informative and insightful illustration of that comes from Netflix, a popular online personalized video-streaming service. For a very long time, the Netflix app interface allowed users to provide feedback in the form of 5-star ratings. Such explicit feedback data was even a part of the famous Netflix Prize competition conducted from 2006 to 2009. Later, in 2017, Netflix replaced ratings with a ``simpler and more intuitive thumbs-up and thumbs-down'' option\footnote{\url{https://about.netflix.com/en/news/goodbye-stars-hello-thumbs}}. Part of the reason was to lessen the confusion about the functionality and purposes of a rating system, and another part was to ``learn even more about your unique tastes''. Recently, however, the binary grading was found to be insufficient and was extended with ``super-like'' or ``double thumb-up''\footnote{\url{https://about.netflix.com/en/news/two-thumbs-up-even-better-recommendations}} to help Netflix learn better representations of user preferences. Essentially, Netflix went from 5-level to 2-level grading system but then had to extend it to at least 3 levels.

From the personalization service perspective, there are two main aspects that emerge from the Netflix's case and seem to be relatively common. One is related to users' competency in properly engaging with the service and understanding its functionality. Another one is related to learning a better recommendation model using the collected behavioral data. Leaving aside the former aspect, we are interested in understanding whether it was necessary to downgrade the rating scale with the aim of learning a more accurate recommendation model. What if there is a better way to deal with the initial 5-star system and gracefully handle rating inconsistencies induced by numerous reasons of a different (and often unknown) nature? We hypothesize that rating values contain useful for recommendations information and reducing noise in feedback by scaling its values can make the signal from ratings even stronger and allows to get better quality of recommendations. In this work, we aim to verify this hypothesis by introducing the notion of ``ratings similarity'' or ``rating proximity''. We will use it to build a smoother representation of user preferences that makes a recommendation model to become more robust against discrepancies in rating patterns. The main contributions of are work are as follows:
\begin{enumerate}
    \item We introduce rating values smoothing in tensor-based model by using similarity matrix.
    \item We considered different similarity matrices and various rating aggregating conditions to explore the influence on the target metric value.
    \item We show that suggested method can outperform strong baselines like matrix factorization methods, tensor-based models and neural autoencoder.
\end{enumerate}

The rest of this paper is organized as follows. 
Section 2 describes the related work. In section 3 we present the problem description, introduce the notion of similarity matrix and explain the core of the proposed method based on the Hybrid SVD \cite{frolov2019hybridsvd} and CoFFee models \cite{frolov2016fifty}. Section 4 gives the overview of the proposed evaluation metrics, namely the MCC metric that is of particular importance for the proposed method since it penalizes for giving irrelevant recommendations. In section 5 we describe the full experimental pipeline: describe the datasets, full data pre-processing, baselines and evaluation steps procedure. In section 6 the obtained results are analyzed. Section 7 concludes the paper.

Source code for reproducing the results are presented in an online repository \footnote{https://github.com/anonymouspap/latte\_recsys}.

\section{Related Work}

As a rule, a standard prediction scheme in recommender systems is based on the prediction of the ratings first. After that the ranking of the items is inferred from the obtained rating predictions. However, this strategy relies on two assumptions. First, that rating is a form of explicit feedback. Second, that the problem of ratings prediction is the same as the optimization of item rankings. A range of papers challenge these assumptions. Thus, MF like models with bias is an attempt to debias individual rating scales. But it will fail in the case of systematic shifts, e.g. when two users have similar tastes but ratings are shifted in opposite directions (one user tends to provide more pessimistic ratings while another user tends to rate more optimistically). Despite the additional complexity, such approaches were shown to be outperformed by much simpler PureSVD model \cite{cremonesi2010performance}.  It was shown to suffer from certain effects related to the fact that ratings are of an ordinal nature, rather then just cardinal values \cite{frolov2016fifty}. Besides, as it was stated in the introduction, some psychological effects like the assimilation/contrast effect can be detected in the form of the historical raing perception. \citet{zhang2017modeling} proposed the Historical Influence Aware Latent Factor Model model that aims to find and mitigate historical distortions in a single user rating. Extracting users' real preferences from the noisy historical ratings solves the problem of distortion both on the macro and micro levels. At the macro-level wrong historical ratings mislead the future buyers and, as a result, it leads to wrong decisions. At the micro-level if the rating doesn't reflect true user's opinion, the recommender system is not able anymore to provide high-quality recommendations. Finally, there are papers that look at the ordinal structure of ratings. For instance, OrdRec model \cite{koren2011ordrec} doesn't treat user feedback values as a number rather than as an order. Such representation allows the users to have different internal scoring scales. Further this idea was supported by Markov Random Fields for
Recommender Systems \cite{liu2016preference} that model user preference relations and are able to discover the second-order and the higher-order interactions among users and items.




\section{Problem Formulation}

\subsection{Problem Description}

While there are many possible inconsistencies in users' perception this work aims to solve the problem of different perception of the rating scale by users. Consider the following example. For some people giving a 5 star rating to an item is an extraordinary event, they do it rarely and only in exceptional cases. Other users are more generous and most of their ratings are 4 and 5 stars. This leads to a situation where the recommender system treats different ratings as different signals which may be misleading. For example, given two users where the first one rated three last films as 3, 4, 4 and another user rated exactly the same films as 4, 5, 5, we should understand that these are the same sets of preferences. To tackle this problem, this work imposes some notion of similarity. 

\subsection{Similarity matrices}

\label{similarity_matrices}

As it was mentioned above, the main advantage of our approach is in the direct modelling of the dependency between values of ratings. The properly modelled nature of the object can boost the performance of prediction in comparison with the more sophisticated (neural) models.

The similarity matrix reflects the proximity between the values of ratings which can be interpreted as a correlation between them. The closer these values are, the larger the correlation between them. However, this dependency can be modelled in different ways. The considered modelling laws are presented below:
\vspace{-0.8ex}
\begin{itemize}
    \item \textit{Linear dependency} : 
    $f(x) = x, x \in [0, 1]$
\item \textit{Sigmoid dependency} : 
$f(x) = \frac{1}{1+e^{-x}}, x \in [-6, 6]$

\item \textit{Arctan dependency} : 
$f(x) = \frac{1}{2}\arctan{x} + \frac{1}{2}, x \in [-\pi / 2, \pi / 2]$
\item \textit{Cube-root dependency} :
$f(x) = \frac{1}{2}\sqrt[3] x +\frac{1}{2}, x \in [-1, 1]$
\end{itemize}

The linear dependency models the linear decrease in correlation between the rating values the far the rating values are from each other. In contrast, the non-linear models (sigmoid, arctan and cubic root) are chosen in order to cluster ratings with low values, ratings with high values and make center ratings far from both of these clusters. Thus, for instance, in case of 5-values rating scale "1" and "2" rating values will be clustered together as well as "4" and "5" rating values. Different non-linear laws will address this pattern to the different extent. For example, correlation between "3" and "2" for the sigmoid law is 0.55 while for the arctan law is 0.67, that is sigmoid law is more <<aggressive>>.

To obtain the specific values in the matrices one can apply the <<law>> in the following way. We choose the range of x values for particular function in a way to make y values be in [0, 1]. Then, the $n$ values $v_1, v_2, ..., v_n$ (where $n$ is the amount of rating values ) are uniformly chosen from this range. Finally, the value in the i-th row j-th column in the correlation matrix is equal to $1-|f(v_i)-f(v_j)|$ where f is the function which models the law of interest.

Below are presented the similarity matrices with different dependencies for the case where ratings are in range from 1 to 5. Correlations are rounded up to two decimal places. The value in the i-th row j-th column in the matrix stands for the correlation between i-th and j-th value of the rating.

\newcommand*{\topbordermatrix}[2]{%
  {\mathop{\bordermatrix{#2}}\limits^{\mbox{#1}}}%
}
\footnotesize
  \begin{align*}
    \topbordermatrix{\textit{(a) Linear dependency}}{ & "1" & "2" & "3" & "4" & "5" \cr
    "1" & 1 & 0.75 & 0.5 & 0.25 & 0 \cr
    "2" & 0.75 & 1 & 0.75 & 0.5 & 0.25 \cr
    "3" & 0.5 & 0.75 & 1 & 0.75 & 0.5 \cr
    "4" & 0.25 & 0.5 & 0.75 & 1 & 0.75 \cr
    "5" & 0 & 0.25 & 0.5 & 0.75 & 1
    }
    &&
    \topbordermatrix{\textit{(b) Sigmoid dependency }}{
     & "1" & "2" & "3" & "4" & "5" \cr
    & 1 & 0.96 & 0.5 & 0.05 & 0 \cr
    & 0.96 & 1 & 0.55 & 0.09 & 0.05 \cr
    & 0.5 & 0.55 & 1 & 0.55 & 0.5 \cr
    & 0.05 & 0.09 & 0.55 & 1 & 0.96 \cr
    & 0 & 0.05 & 0.5 & 0.96 & 1 \cr
    }
    &&
    \topbordermatrix{\textit{(c) Arctan dependency}}{
     & "1" & "2" & "3" & "4" & "5" \cr
    & 1 & 0.83 & 0.5 & 0.17 & 0 \cr
    &  0.83 & 1 & 0.67 & 0.33 & 0.17 \cr
    & 0.5 & 0.67 & 1 & 0.67 & 0.5 \cr
    & 0.17 & 0.33 & 0.67 & 1 & 0.83 \cr
    & 0 & 0.17 & 0.5 & 0.83 & 1 
    }
    &&
    \topbordermatrix{\textit{(d) Cube-root dependency}}{ 
     & "1" & "2" & "3" & "4" & "5" \cr
    & 1 & 0.9 & 0.5 & 0.1 & 0 \cr
    & 0.9 & 1 & 0.6 & 0.21 & 0.1 \cr
    & 0.5 & 0.6 & 1 & 0.6 & 0.5 \cr
    & 0.1 & 0.21 & 0.6 & 1 & 0.9 \cr
    & 0 & 0.1 & 0.5 & 0.9 & 1
    }
  \end{align*}
\normalsize

\subsection{Models}

\subsubsection{Prerequisites}

The idea of the proposed approach lies at the intersection of the HybridSVD model \cite{frolov2019hybridsvd} and 
CoFFee model (Collaborative Full Feedback Model) \cite{frolov2016fifty}.

The novelty of the HybridSVD model \cite{frolov2019hybridsvd} was in taking into consideration the side similarity between users and/or between items. The standard PureSVD approach can be reformulated as an eigendecomposition problem of a scaled user-based or item-based cosine similarity. In the user-based case the eigendecomposition is made for $RR^T$ matrix, where $R \in \mathbb{R}^{M \times N}$ is a matrix of interactions with M users and N items. If we take the $k_{ij} = r^T_{i}r_j$ element of this matrix, that is the similarity score between users i and j, the contribution of each item is taken into consideration only in case it is presented in the preference profiles of both users i and j. So, such problem statement doesn't take into account the possible dependency between items (and users). To tackle this problem, the authors suggest to replace scalar product $r^T_{i}r_j$ with a bilinear form $r^T_{i}Sr_j$, where $S \in \mathbb{R}^{N \times N}$ is a symmetric matrix that incorporates side information between items.

The second model that inspired this paper is the CoFFee model \cite{frolov2016fifty}. According to a well-known folding-in technique \cite{ekstrand2011collaborative} the prediction of ratings in a PureSVD case for the new users can be obtained as $VV^Tp$, where p is a vector of preferences of a user and V is an orthogonal factor of latent movies representation obtained in the standard SVD approach. Now assume the extreme case where we have only one rating provided by the user. Irrespective of the value of this rating (either it will be 2 or 5) it will just rescale the values in the product $VV^T$. Since the final ranking doesn't depend on the value of this scaling factor in this particular case, it highlights the problem of the suggested method. For solving it, the authors propose the following framework. 

This model predicts the relevance scores which can be further used for selecting top-n recommendations:
\[
f_U: User \times Item \times Rating \rightarrow Relevance \hspace{0.5ex} Score
\]

Let's have a tensor $\mathbb{X} \in \mathbb{R^{M\times N \times K}}$, where M is the total amount of users, N is the total amount of items and K is the total amount of rating values. The values of the tensor $\mathbb{X}$ are binary:
\begin{equation}
\begin{cases} 
x_{ijk} = 1 \ \text{if (i,j,k)} \in H\\
x_{ijk} = 0 \ \text{otherwise}  
\end{cases}
\end{equation}

where $H$ is a history of recommendations. So, triplet (i, j, k) characterizes whether user \textit{i} has seen movie \textit{j} and gave it rating \textit{k}. Next step is the tensor decomposition for obtaining latent representations of users, movies and ratings. Using Tucker decomposition \cite{kolda2009tensor} (which authors proved to be the most efficient) one can obtain the following representation:
\begin{equation}
\label{eq:tensor10}
 \mathbb{X} \approx \mathbb{G} \times_1 \mathbb{U} \times_2 \mathbb{V} \times_3 \mathbb{W} 
\end{equation}

In equation (\ref{eq:tensor10}) $\mathbb{G} \in \mathbb{R}^{r_1\times r_2 \times r_3}$, 
$U \in \mathbb{R}^{M\times r_1}$,
$V \in \mathbb{R}^{N\times r_2}$,
$W \in \mathbb{R}^{K\times r_3}$, where $(r_1, r_2, r_3)$ is a multilinear rank of
the decomposition.

This decomposition can be effectively
computed with a higher order orthogonal iterations (HOOI) algorithm \cite{de2000best}.

Using matrices obtained from decomposition predicted ratings for a user $i$ given the binary matrix of preferences $P_i$ can be calculated as:
\begin{equation}
\label{eq:coffee}
R_i \approx VV^TP_iWW^T
\end{equation}

where $R_i \in \mathbb{R^{N\times K}}$, $P_i \in \mathbb{R^{N\times K}}$. In contrast to the conventional models that predict a single score for an item, the output of this model is a vector of scores where each score states for the likeliness of each rating to belong to an item. These scores metaphorically compared by  \citet{frolov2016fifty} with <<fifty shades of ratings>> allow to better represent ratings prediction structure and can enforce ranking by aggregating this information to the different extent.

\subsubsection{LaTTe}

Our model which we call LaTTe (Latent Attention Model) combines inside itself the idea of tensor representation and similarity matrix between ratings. Idea of similarity matrix is taken from the HybridSVD model while the tensor representation of triples user-item-rating relates to the CoFFee model. It was proved by \citet{burke2002hybrid} that models that combine both collaborative data and side information demonstrate better performance in many cases. The exact similarity matrices are taken from section \nameref{similarity_matrices}. 

Following \citet{frolov2018revealing} the solution of the eigendecomposition problem can be represented in the form of standard
truncated SVD of an auxiliary matrix and \label{eq:tensor} is approximated as:

\begin{equation}
\label{eq:tensor2}
 \widehat{\mathbb{X}} \approx \mathbb{G} \times_1 \widehat{\mathbb{U}} \times_2 \widehat{\mathbb{V}} \times_3 \widehat{\mathbb{W}} 
\end{equation}

Note that 
$\widehat{W} \in \mathbb{R}^{K\times r_3}$ 
corresponds to the auxiliary latent space. The latent representation of ratings in the original space is then represented as:

\begin{equation}
\label{eq:orig_latent}
  W = K^{-\frac{1}{2}}\widehat{W}
\end{equation}

From the HybridSVD model and folding-in for the CoFFee model it's easy to obtain the folding-in for LaTTe:

\begin{equation}
R_i \approx VV^TP_iK^{\frac{1}{2}}WW^TK^{-\frac{1}{2}}
\end{equation}





\section{Evaluation metrics}
\label{metrics}

As it was described in \citet{frolov2016fifty} standard metrics of offline evaluation are insensitive to irrelevant items predictions. We overestimate the performance of our model when we use metrics that don't take into account the rating left by user as a feedback. For example, a user actually watches the movie we recommended but put a dislike -- this situation shouldn't be considered as a successful recommendation.

One of the solutions to this problem is to divide rating values into two classes using a negativity threshold value\footnote{values of ratings which are less that threshold are considered as negative, and vice versa} \cite{frolov2016fifty}. Using this notion we can calculate true positive (TP), false positive (FP), true negative (TN), false negative (FN). It should be noted that we don't take into account those recommendations for which we didn't get any feedback from a user. It can't be interpreted as false positive since we don`t know the real rating -- it may be the case that the user just missed our recommendation and still has a chance to take it later providing positive or negative feedback.

Standard metrics like precision and recall now can be calculated in a usual way. On the contrary, HR and MRR metrics \cite{deshpande2004item} need some changes: we can still calculate them but we should divide each of them into two parts (HR for positive ratings, HR$^+$, and HR for negative ratings, HR$^-$; MRR for positive ratings, MRR$^+$ and MRR for negative ratings, MRR$^-$). Both HR and MRR show how many items are relevant for users -- that's why we would like them to be as high as possible. But this logic changes for negative ratings -- they are not relevant for users. So, HR$^-$ and MRR$^-$ are aimed to be as low as possible.

This approach allows us to track both aspects of our model's performance. Namely, predicting relevant items and avoiding irrelevant ones. This makes our task quite close to a classification task -- more relevant metrics are available, e.g. ROC-AUC, PR-AUC, etc. \cite{frolov2016fifty}.

\begin{figure}[h]
\includegraphics[width=8cm]{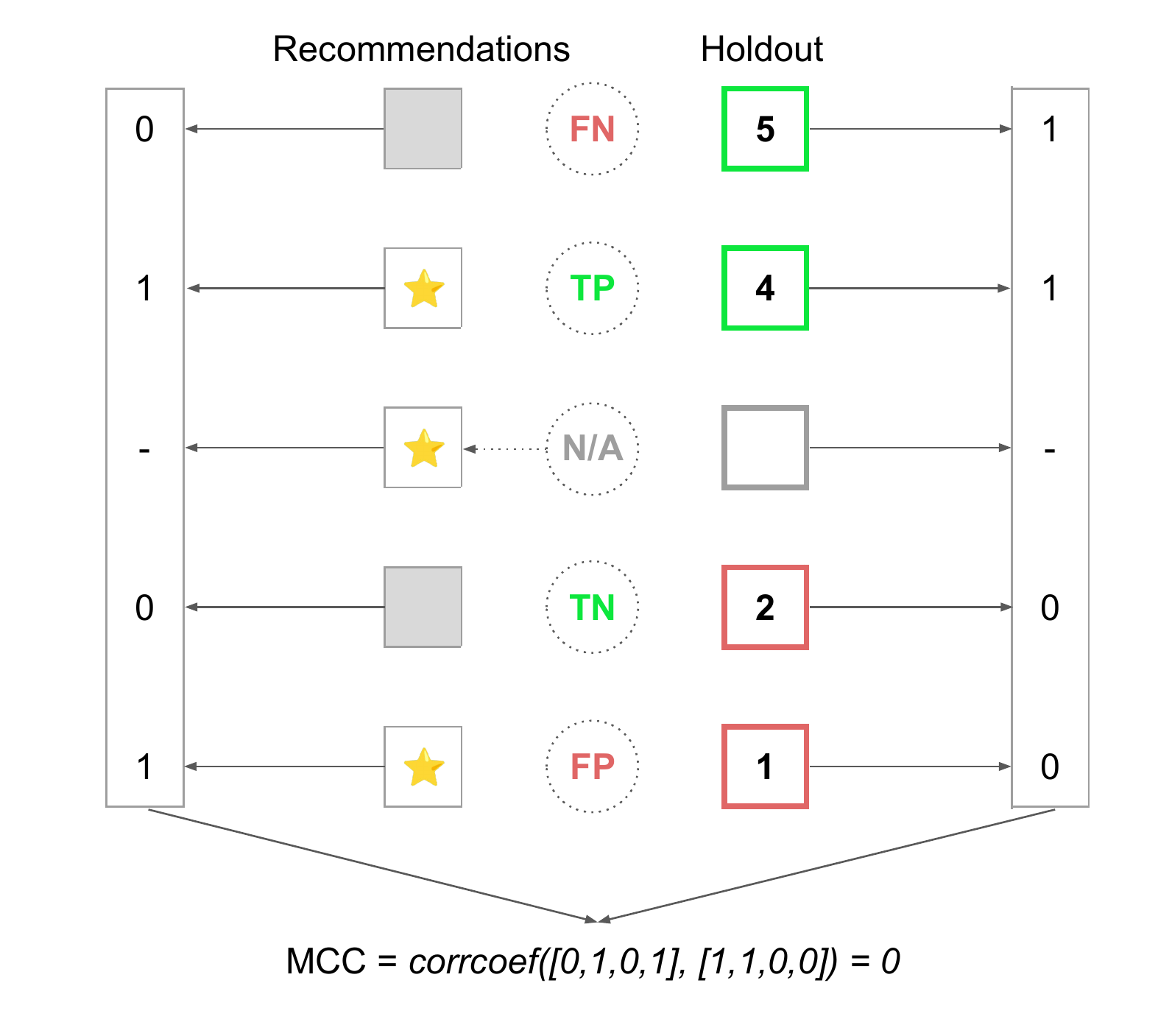}
\caption{The graph shows an example of calculating MCC metric. Stars in the recommendations column show recommended items. Numbers in the holdout column show actual rating for predicted items (if they are in holdout, otherwise there is no rating). Numbers that highlighted by green color are "positive" ratings, by red color -- "negative" ratings. Both columns can be converted to one-hot vectors where 1 corresponds to positive class, 0 -- to negative class. These vectors can be used to calculate MCC if we find correlation between them}
\label{fig:metrics}
\centering
\end{figure}

However, these metrics may show different results, i.e. one model is good for a positive class, another one is good for a negative class. We want to have one metric to see the overall performance of the model and tune it accurately. So, we need to combine results for both classes in one metric. For that we use \textit{Matthews correlation coefficient} -- another classification metric.
This coefficient shows correlation between relevant and predicted items\footnote{between true and predicted labels in terms of binary classification}. So we want this coefficient to be as high as possible. You can find example of correlation calculations in Figure \ref{fig:metrics} -- one may notice that unknown rating is not included due to reasons described earlier.

We can also derive Matthews correlation coefficient in terms of TP, FP, TN, FN:
$$
MCC = \frac{TP \cdot TN - FP \cdot FN}{\sqrt{(TP+FN) \cdot (TP + FP) \cdot (TN + FP) \cdot (TN + FN)}}
$$

For our example (see Figure \ref{fig:metrics}): $MCC = \frac{1 \cdot 1 - 1 \cdot 1}{\sqrt{(1 + 1) \cdot (1 + 1) \cdot (1 + 1) \cdot (1 + 1)}} = 0$

This metric combines classical precision and recall (HR in our case) in one metric \cite{schroder2011setting}.
Additionally, this metric is insensitive to class imbalance.

We also consider coverage (the amount of unique items recommended to all users divided by the amount of unique items in the training dataset) to estimate the diversity of predictions.

Totally, we consider the following metrics to evaluate models: HR$^+$, HR$^-$, MRR$^+$, MRR$^-$, coverage, MCC. MCC is the metric we use to tune models.



\section{Experimental setup}
\subsection{Datasets}
\begin{figure}[h]
\includegraphics[width=14cm]{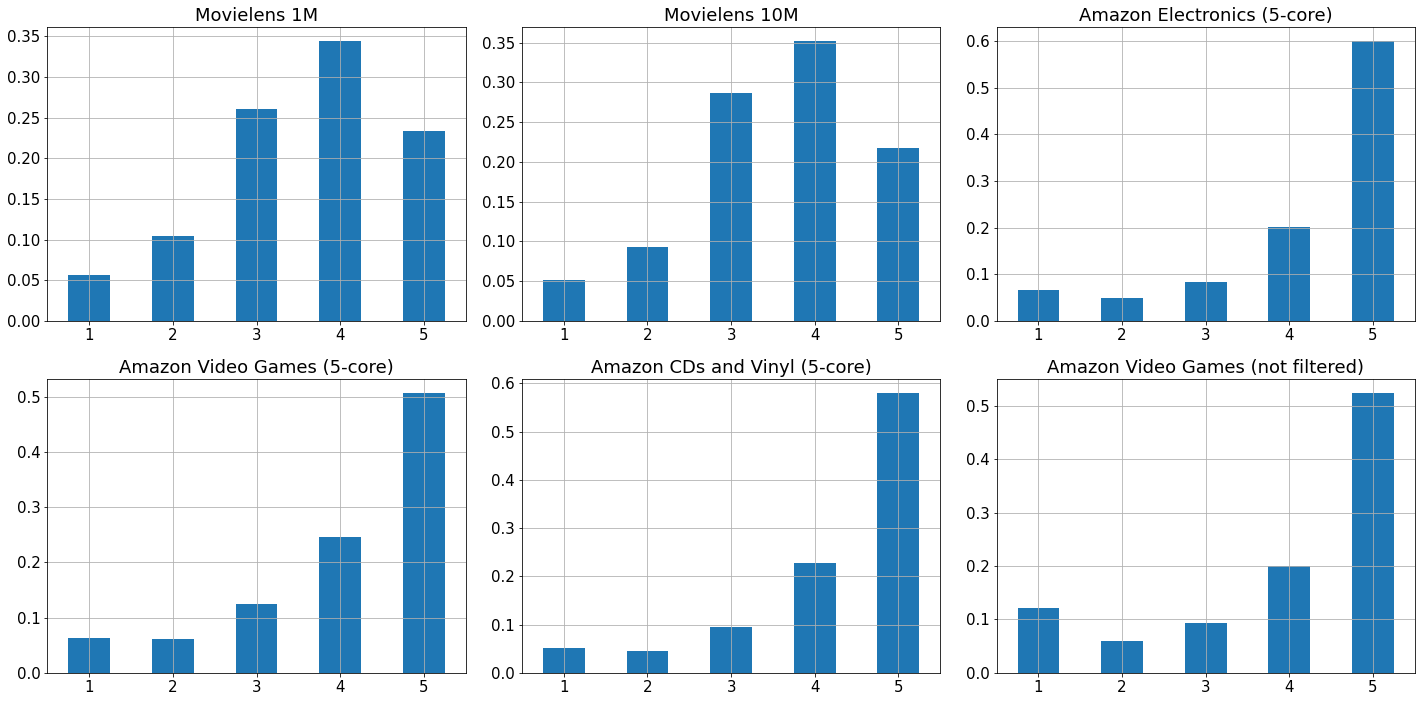}
\caption{The distribution of rating values.}
\label{fig:hists}
\centering
\end{figure}

To test our hypothesis we use publicly available Movielens \footnote{\url{https://grouplens.org/datasets/movielens/}} (1M and 10M) and Amazon \footnote{\url{http://snap.stanford.edu/data/amazon/productGraph/categoryFiles/}} "CDs and Vinyl", "Electronics" and "Video Games" datasets. Each of these Amazon datasets is the subset of the original data in which all users and items have at least 5 reviews (it is the so-called 5-core filtering). Also we decided to take one not filtered Amazon dataset - "Video Games" from Amazon Review Data (2018) \footnote{\url{https://nijianmo.github.io/amazon/index.html}}. Data description of each dataset is presented in Table \ref{tab:datasets}. Next, we analyzed the distributions of rating values in all of the suggested datasets. These distributions are presented in Figure \ref{fig:hists}. For Movielens 10M we transformed ratings into 5 -scale range (more details in section 6.3 \nameref{algorithms}). As you can see, the distributions of ratings in Amazon dataset are very biased towards "5" value. The different pattern can be observed in the Movielens datasets. Distributions of these datasets look like shifted normal distribution. 

\begin{table}[h]
\caption{Numerical statistics of the datasets.}
\label{tab:datasets}

\begin{tabular}{ccccc}
\hline
Dataset                           & N\_review & N\_users & N\_items & N\_rates \\ \hline
Movielens 1M                      & 1000209   & 6040     & 3706     & 5        \\
Movielens 10M                     & 10000054  & 69878    & 10677    & 10       \\
Amazon CDs and Vinil (5-core)     & 1097592   & 75258    & 64443    & 5        \\
Amazon Electronics (5-core)       & 1689188   & 192403   & 63001    & 5        \\
Amazon Video Games (5-core)       & 231780    & 24303    & 10672    & 5        \\
Amazon Video Games (not filtered) & 1324753   & 826767   & 50210    & 5        \\ \hline
\end{tabular}
\end{table}

\subsection{Data preprocessing}

In order to simulate experiments as similar as possible to the real life process, we conducted special preprocessing of data by the timestamp parameter. First of all, we decided to split data on train and test parts -- last 20\% of the most recent reviews (by timestamp) were included to the test part. An exception was made for Amazon "Electronics" dataset where we splitted data using 5\% of the most recent reviews to make holdout size be equal to 10k (the size that doesn't overload computations and is still efficient for making consistent conclusions). This split was made to avoid "predictions from future". Then, the test part of data was split into validation and test part by using "leave-last-out" method of obtaining holdout items twice: last items of all users from test data went to test holdout, and last items from remaining test data went to validation holdout. Also we removed items from both of the holdouts that were not presented in the training dataset. 

\subsection{Algorithms}

\label{algorithms}

We compared the suggested method with existing algorithms for generating top-n recommendations. In this comparison were used the following algorithms:
\begin{itemize}
    \item \textit{Random} model that generates recommendations randomly for all users.
    \item \textit{Most Popular (MP)} model always predicts the most popular items in the dataset for all users.
    \item \textit{Normalized PureSVD} - EigenRec-inspired \cite{frolov2019hybridsvd, nikolakopoulos2019eigenrec}  model uses singular value decomposition (SVD) of the user-item matrix to obtain latent vectors that are used in the top-n generating method. Additionally it reweights "popularity" of items by multiplication on diagonal matrix that allows to avoid predicting only popular items. This baseline is stronger than usual PureSVD.
    \item \textit{EASEr} \cite{steck2019embarrassingly} is linear model with constraints for final weights matrix.
    \item \textit{MultiVAE} \cite{liang2018variational} is neural network approach, which is a regularized Variational Autoencoder with multinomial likelihood, denoted as Mult-VAE ${ }^{P R}$.
    \item \textit{CoFFee} \cite{frolov2016fifty} model uses tensor factorization algorithm (HOOI) to generate recommendations using 3-dimensional representation of input user-item matrix. In this model the normalization matrix were also used to reweight "popularity" of items.
    \item \textit{LaTTe} is the proposed tensor-based approach with ``smoothed'' ratings. This model uses normalization as well as MP and CoFFee models.
\end{itemize}

To make a valid comparison of the models we tune them. 
The first two baselines haven't got any hyperparameters for tuning. In Normalized PureSVD model we tuned two parameters: normalization factor that was used in the normalization matrix's power and dimensionality (rank) of the latent space of the users and items in SVD. First parameter took values from $[0, 2]$ with step $0.1$ and for second one we use the following grid - $(2 \times 2^i, 3 \times 2^i)$, $i \in [5, 8]$. In EASEr model we tuned L2-norm regularization parameter $\lambda$ from [50, 950] with step 50. 

When it came to the tuning of Mult-VAE ${ }^{P R}$ we studied tuning process suggested in the original paper \cite{liang2018variational}. During it we focused on three hyperparameters: annealing factor $\beta$ (a coefficient in front of the Kullback-Leiber term, which is a part of the final loss), network architecture (number of neurons in the 1st and 2nd hidden layers) and batch size. As it comes to the annealing factor we followed the original idea with early stopping: once $\beta$ achieved the best Matthew's correlation coefficient (MCC) for the last 10 epochs, we fixed it and trained the network from the scratch. After tuning we picked one, which achieved the best MCC for training. In addition, we checked different architectures for Mult-VAE${ }^{P R}$ model because of the poor description of the original tuning parameters of hidden layers. For all such experiments we used batch size of 500 as it was suggested in the original paper. 

Other tensor-based models (CoFFee and LaTTe) are the extension of Normalized PureSVD model in 3D space, so we consider to use normalization factor from PureSVD in these models, intuitively because "popularity" of the items doesn't change in the extended 3D space. The latent space dimensionalities were tuned for the both of the tensor-factorization models. Besides, we consider to take the equal dimensionality for users and items latent spaces, so these  spaces' dimensionality numbers were taken from the same grid as it was used in the Normalized PureSVD tuning. The ratings' rank number was taken from [2, 5] with step 1. Additionally, we made hyperparameters' tuning of different LaTTe models with all our proposed similarity matrices.

\subsection{Evaluation process}

Metrics described in section \ref{metrics} were calculated for top-10 items during the evaluation process. As our target metric we consider MCC@10.

Since ratings in tensor-based models (CoFFee and LaTTe) have their own dimension, these models predict \emph{the vector of scores for each pair user-item}, instead of just a single number like in the conventional matrix factorization models. Each component of this vector describes a score for this pair based on the condition of a possible user's feedback to the item. 

To deal with the different ratings' range we propose universal method of taking into account ratings influence.
We convert all ranges of ratings into 5-values scale by considering top 20\% of the sorted in descending order ratings' values as "5", values from 20-40\% we considered as "4" and so on.
E.g. in 10-values scale scores of "9" and "10" in the vector are summed up into one value which we assign to rating "5" in the new scale.
After such transformation we get the vector with only 5 values.

However, this rating scale transformation method can be changed for different systems if there is some side information about rating values' perception by users inside the system.
Also we make an assumption that rating "3" and higher can be treated as positive rating (it is needed to calculate TP, TN, FP and FN).

We combined predicted scores using its hidden "context" described above as it was suggested in \cite{frolov2016fifty}. In our experiments we consider several combinations of elements of this vector: only "5" ratings, $ "4" + "5"$, $"3" + "4" + "5"$ and $"3"+"4"+"5"-"1"-"2"$ "contexts". The last expression means that we sum the last three scores of the predicted vector and subtract scores for values "1" and "2" to obtain the final score for the particular pair user-item. The intuition is that in the last setup we encourage the high scores for the high rating values (3, 4, 5) and additionally penalize high scores for the low values of ratings (1,2).

\section{Results}


The results for models (tuned on the valudation data) on the test datasets are presented in Table \ref{tab:results}.

For Movielens datasets our model shows the best $MCC@10$ value among all other models and relatively high $HR^+$ value. We concluded that such results were achieved because of the close connectivity between suggested similarity matrices and the distribution of ratings on these datasets.

On the Amazon 5-core filtered datasets our model shows the same or lower performance in comparison with other models. Distributions of ratings on these datasets (Figure \ref{fig:hists}) can be the reason of such results. Our approach of "smoothing" ratings assumed a close similarity between "5" and "4" rating values and greater differences between "3" and "4" values. However, this property stays unsatisfied according to the plotted histogram of ratings. Besides, the choice of the threshold for defining positive ratings can be the reason of the worse result on the datasets too. Due to this observation the results could have been possibly improved in case of a different similarity matrix. For instance, choosing a matrix that creates a considerable gap between "5" value rating and another values in terms of correlation could have resolved the issue. Beyond this, choosing "5" as a new threshold for positive ratings could have been a better strategy.

For non-filtered Amazon dataset for video games we get the best result with considerable gap from other models. We can see a little improvement in performance of LaTTe compare to unfiltered version of this dataset. While for the PureSVD (which was the best on filtered version) there is a decrease in performance by 2\%. One of the possible reasons of such situation is that our model is more robust to the noise we have in data: users or items with small number of available reviews have not much useful information we can include in the model. However, our approach allows to deal with this problem since we consider also correlation between ratings which give us more data in comparison to other models.

\begin{table}[h]
\caption{Comparison of the models' performance on each of the datasets.}
\label{tab:results}
\begin{tabular}{ccccccccc}
\hline
Dataset & Metric@10 & Random & MP & PureSVD & EASEr & MultiVAE & CoFFee & LaTTe \\
\hline
\multirow{6}{30mm}{Movielens 1m} & HR$^+$ & 0.0023 & 0.0224 & 0.0495 & 0.0558 & 0.0587 & 0.0431 & \textbf{0.0598} \\
& HR$^-$ & 0.0000 & 0.0006 & 0.0058 & 0.0040 & 0.0052 & 0.0023 & 0.0029 \\
& MRR$^+$ & 0.0012 & 0.0078 & 0.0198 & \textbf{0.0225} & 0.0188 & 0.0166 & 0.0196 \\
& MRR$^-$  & 0.0000  & 0.0001  & 0.0010 & 0.0011          & 0.0015          & 0.0005 & 0.0005          \\
& Coverage  & 0.9936  & 0.0301  & 0.1302          & 0.2713          & \textbf{0.3344} & 0.2354 & 0.2097          \\
& MCC       & 0.0221  & 0.0604  & 0.0447          & 0.0711          & 0.0641          & 0.0711 & \textbf{0.0875} \\
\hline
\multirow{6}{30mm}{Movielens 10m}                     & HR$^+$   & 0.0009  & 0.0425  & \textbf{0.0920} & 0.0902          & 0.0836          & 0.0671 & 0.0808          \\
                                                   & HR$^-$   & 0.0000  & 0.0050  & 0.0056          & 0.0055          & 0.0061          & 0.0035 & 0.0043          \\
                                                   & MRR$^+$  & 0.0002  & 0.0160  & \textbf{0.0370} & 0.0366          & 0.0305          & 0.0279 & 0.0325          \\
                                                   & MRR$^-$  & 0.0000  & 0.0016  & 0.0019          & 0.0020          & 0.0020          & 0.0014 & 0.0016          \\
                                                   & Coverage  & 1.0000  & 0.0276  & 0.1646          & 0.2144          & \textbf{0.3491} & 0.1091 & 0.1759          \\
                                                   & MCC       & 0.0120  & 0.0212  & 0.0769          & 0.0757          & 0.0635          & 0.0706 & \textbf{0.0771} \\ \hline
\multirow{6}{30mm}{Amazon CDs\_and\_Vinyl (5-core)}   & HR$^+$   & 0.0001  & 0.0027  & 0.0268          & \textbf{0.0654} & 0.0394          & 0.0186 & 0.0185          \\
                                                   & HR$^-$   & 0.0000  & 0.0001  & 0.0012          & 0.0020          & 0.0013          & 0.0005 & 0.0003          \\
                                                   & MRR$^+$  & 0.0000  & 0.0007  & 0.0122          & \textbf{0.0320} & 0.0173          & 0.0080 & 0.0077          \\
                                                   & MRR$^-$  & 0.0000  & 0.0001  & 0.0005          & 0.0009          & 0.0006          & 0.0002 & 0.0001          \\
                                                   & Coverage  & 0.9703  & 0.0003  & 0.0680          & \textbf{0.5785} & 0.4853          & 0.0980 & 0.0414          \\
                                                   & MCC       & 0.0027  & -0.0019 & 0.0008          & 0.0194          & 0.0122          & 0.0128 & \textbf{0.0196} \\ \hline
\multirow{6}{30mm}{Amazon Electronics (5-core)}       & HR$^+$   & 0.0002  & 0.0219  & 0.0295          & \textbf{0.0522} & 0.0455          & 0.0223 & 0.0221          \\
                                                   & HR$^-$   & 0.0000  & 0.0010  & 0.0010          & 0.0028          & 0.0021          & 0.0006 & 0.0009          \\
                                                   & MRR$^+$  & 0.0000  & 0.0103  & 0.0125          & \textbf{0.0288} & 0.0216          & 0.0111 & 0.0119          \\
                                                   & MRR$^-$  & 0.0000  & 0.0005  & 0.0003          & 0.0014          & 0.0008          & 0.0003 & 0.0003          \\
                                                   & Coverage  & 0.9387  & 0.0002  & 0.0092          & \textbf{0.4382} & 0.2968          & 0.0237 & 0.0084          \\
                                                   & MCC       & 0.0049  & 0.0254  & 0.0374          & 0.0350          & \textbf{0.0383} & 0.0347 & 0.0295          \\ \hline
\multirow{6}{30mm}{Amazon Video Games (5-core)}       & HR$^+$   & 0.0009  & 0.0100  & 0.0739          & \textbf{0.0840} & 0.0730          & 0.0485 & 0.0652          \\
                                                   & HR$^-$   & 0.0002  & 0.0013  & 0.0035          & 0.0057          & 0.0060          & 0.0018 & 0.0037          \\
                                                   & MRR$^+$  & 0.0003  & 0.0033  & 0.0339          & \textbf{0.0375} & 0.0288          & 0.0221 & 0.0285          \\
                                                   & MRR$^-$  & 0.0002  & 0.0005  & 0.0018          & 0.0027          & 0.0026          & 0.0011 & 0.0018          \\
                                                   & Coverage  & 0.9965  & 0.0013  & 0.3458          & \textbf{0.6076} & 0.5472          & 0.1002 & 0.1422          \\
                                                   & MCC       & -0.0076 & -0.0051 & \textbf{0.0522} & 0.0373          & 0.0220          & 0.0482 & 0.0415          \\ \hline
\multirow{6}{30mm}{Amazon Video Games (not filtered)} & HR$^+$   & 0.0001  & 0.0350  & 0.0450          & 0.1233          & \textbf{0.1331} & 0.0185 & 0.0368          \\
                                                   & HR$^-$  & 0.000   & 0.0030  & 0.0035          & 0.0117          & 0.0147          & 0.0012 & 0.0016          \\
                                                   & MRR$^+$  & 0.000   & 0.0108  & 0.0268          & 0.0636          & \textbf{0.0646} & 0.0093 & 0.0208          \\
                                                   & MRR$^-$  & 0.000   & 0.0010  & 0.0018          & 0.0056          & 0.0066          & 0.0005 & 0.0007          \\
                                                   & Coverage  & 0.9984  & 0.0003  & 0.0266          & \textbf{0.3776} & 0.1993          & 0.0094 & 0.0121          \\
                                                   & MCC       & 0.0043  & 0.0239  & 0.0318          & 0.0368          & 0.0222          & 0.0254 & \textbf{0.0469} \\ \hline
\end{tabular}
\end{table}

Also to show the importance of choice of the right approach for scaling ratings we compare values of target metric $MCC@10$ that were obtained by LaTTe model with different similarity matrices on all datasets. Results of the comparison are presented in Table \ref{tab:results_matrices}. We concluded that for each dataset the best value of target metric was reached with different similarity matrix and with different context aggregation condition. To provide more intuition, the unique combination between context aggregation and proximity matrix can boost the performance considerably. For instance, for Movielens 10M the best were "4"+"5" condition and sigmoid law while for the Amazon Video Games (not filtered) dataset "3"+"4"+"5"-"2"-"1" condition and cube-root law turned out to outperform other combinations. Beyond this, one more important observation can be made. Since for some of the datasets only one context aggregation value is prevalent irrespective of the law it means that changing negativity threshold (which in our case is "3") may lead to better results. If there is no signal from current negativity threshold then choosing another threshold value can increase the performance.

Also we noticed that on some of the Amazon datasets we obtained the much higher MCC@10 value on test data by LaTTe models tuned by HR@10. For example, on Video Games (5-core) dataset we reached 0.0480 value of MCC@10 on test data on LaTTe model tuned by HR@10. This value is much better than performance of the model tuned by target metric. Moreover, LaTTe model shows worse result in comparison with CoFFee model on this dataset. It means that incorrect choice of scaling ratings technique can not only slightly decrease the quality, but can make it even worse by "confusing" the model. In particular, according to the MCC metrics we try to increase the amount of good recommendations in line with the decrease of the amount of bad recommendations. In case the best context aggregation threshold is defined in a wrong way it misleads the model in a way it can't properly distinguish what is bad and good rating value. 

For example, if we include "3" rating value in the threshold for defining positive ratings while it's not true we start recommending the items with this rating in the case of MCC tuning. However, tuning by HR allows to make recommendations irrespective of the defined threshold and, as a consequence, it leads to better achieved results by both metrics.

To sum up, smoothing of ratings lead to better results for some datasets. Choice of matrices and negativity threshold strongly depends on the data. Using side information about users' feedback may also be useful. 

\begin{table}[h]
\caption{Comparison of the MCC@10 metric on all datasets obtained by our method on different similarity matrices. Also for each  metric value we added the best context aggregation condition.}
\label{tab:results_matrices}
\begin{tabular}{ccccc}
\hline
Dataset                           & Linear                                                         & Sigmoid                                                                        & Arctan                                                                & Cube-root                                                                     \\ \hline
Movielens 1M                      & \textbf{\begin{tabular}[c]{@{}c@{}}0.0875 \\ "5"\end{tabular}} & \begin{tabular}[c]{@{}c@{}}0.0816\\  "5"\end{tabular}                          & \begin{tabular}[c]{@{}c@{}}0.0810 \\ "5"\end{tabular}                 & \begin{tabular}[c]{@{}c@{}}0.0871\\ "3"+"4"+"5"-"2"-"1"\end{tabular}          \\ \hline
Movielens 10M                     & \begin{tabular}[c]{@{}c@{}}0.0644 \\ "4"+"5"\end{tabular}      & \textbf{\begin{tabular}[c]{@{}c@{}}0.0771\\  "4" + "5"\end{tabular}}           & \begin{tabular}[c]{@{}c@{}}0.0646\\  "4"+"5"\end{tabular}             & \begin{tabular}[c]{@{}c@{}}0.0748 \\ "4"+"5"\end{tabular}                     \\ \hline
Amazon CDs and Vinil (5-core)     & \begin{tabular}[c]{@{}c@{}}0.0124 \\ "5"\end{tabular}          & \textbf{\begin{tabular}[c]{@{}c@{}}0.0196 \\ "3"+"4"+"5"-"2"-"1"\end{tabular}} & \begin{tabular}[c]{@{}c@{}}0.0164 \\ "3"+"4"+"5"-"2"-"1"\end{tabular} & \begin{tabular}[c]{@{}c@{}}0.0148 \\ "5"\end{tabular}                         \\ \hline
Amazon Electronics (5-core)       & \begin{tabular}[c]{@{}c@{}}0.0282\\ "4"+"5"\end{tabular}       & \textbf{\begin{tabular}[c]{@{}c@{}}0.0295\\ "4" + "5"\end{tabular}}            & \begin{tabular}[c]{@{}c@{}}0.0281\\ "5"\end{tabular}                  & \begin{tabular}[c]{@{}c@{}}0.0276\\ "3"+"4"+"5"-"2"-"1"\end{tabular}          \\ \hline
Amazon Video Games (5-core)       & \textbf{\begin{tabular}[c]{@{}c@{}}0.0415\\ "5"\end{tabular}}  & \begin{tabular}[c]{@{}c@{}}0.0382\\ "5"\end{tabular}                           & \begin{tabular}[c]{@{}c@{}}0.0367\\ "5"\end{tabular}                  & \begin{tabular}[c]{@{}c@{}}0.0382\\ "5"\end{tabular}                          \\ \hline
Amazon Video Games (not filtered) & \begin{tabular}[c]{@{}c@{}}0.0373\\ "5"\end{tabular}           & \begin{tabular}[c]{@{}c@{}}0.0435\\ "5"\end{tabular}                           & \begin{tabular}[c]{@{}c@{}}0.0389\\ "5"\end{tabular}                  & \textbf{\begin{tabular}[c]{@{}c@{}}0.0469\\ "3"+"4"+"5"-"2"-"1"\end{tabular}} \\ \hline
\end{tabular}
\end{table}

\section{Conclusions}

The proposed approach shows that the appropriate choice of ratings' "smoothing" (similarity matrix) can considerably improve performance of the proposed recommendations. We observe this effect due to the decreasing discrepancy in users' rating perception. This confirms our hypothesis about powerful signal from ratings we can get when we reduce the noise. Besides, better ratings representation can lead to a better quality of recommendations without the need of significantly increasing the complexity of solution.

We have shown that integration of specific similarity matrix in existed model (CoFFee in our case) increases the quality of recommendations in comparison to the original model -- we can see more relevant items and/or less irrelevant ones. For some datasets we achieved the best result even among all considered models (including EASEr and MultiVAE). 

Our approach can be potentially applied to other methods as well and is not specific to tensor factorization. Extending it to neural or other approaches is an interesting research direction.

However, the choice of this smoothing should be based on the initial nature of the dependency between rating values. Revealing this dependency is a kind of art. As it has been showed, one of the hints can be found in the distributions of ratings.

\bibliographystyle{ACM-Reference-Format}

\newpage

\bibliography{main}




\end{document}